\documentclass[preprint2,longabstract]{aastex}

\usepackage{natbib}

\slugcomment{Accepted for publication in ApJ}

\shorttitle{Infrared and kinematic properties of G\,196--3\,B}
\shortauthors{Zapatero Osorio et al$.$}

\begin{document}

\title{Infrared and kinematic properties of the substellar object G\,196--3\,B}

\author{M. R. Zapatero Osorio}
\affil{Centro de Astrobiolog\'\i a (CSIC-INTA). Crta. Ajalvir km 4. E-28850 Torrej\'on de Ardoz, Madrid, Spain}
\email{mosorio@cab.inta-csic.es}
\author{R. Rebolo\altaffilmark{1,2}, G. Bihain\altaffilmark{1,2}, V. J. S. B\'ejar\altaffilmark{1}}
\affil{Instituto de Astrof\'\i sica de Canarias (IAC). C/. V\'\i a L\'actea s/n. E-38205 La Laguna, Tenerife, Spain}
\email{rrl@iac.es,gbihain@iac.es,vbejar@iac.es}
\author{J. A. Caballero}
\affil{Centro de Astrobiolog\'\i a (CSIC-INTA). Crta. Ajalvir km 4. E-28850 Torrej\'on de Ardoz, Madrid, Spain}
\email{caballero@astrax.fis.ucm.es}
\and
\author{C. \'Alvarez}
\affil{Instituto de Astrof\'\i sica de Canarias (IAC). C/. V\'\i a L\'actea s/n. E-38205 La Laguna, Tenerife, Spain}
\email{carlos.alvarez@gtc.iac.es}

\altaffiltext{1}{Departamento de Astrof\'\i sica, Universidad de La Laguna, Tenerife, Spain}
\altaffiltext{2}{Also at Consejo Superior de Investigaciones Cient\'\i ficas (CSIC), Spain}

\begin{abstract}
We report unusual near- and mid-infrared photometric properties of \object{G\,196--3\,B}, the young substellar companion at 16\arcsec~from the active M2.5-type star \object{G\,196--3\,A}, using data taken with the IRAC and MIPS instruments onboard {\sl Spitzer}. \object{G\,196--3\,B} shows markedly redder colors at all wavelengths from 1.6 up to 24 $\micron$ than expected for its spectral type, which is determined at L3 from optical and near-infrared spectra. We discuss various physical scenarios to account for its reddish nature, and conclude that a low-gravity atmosphere with enshrouded upper atmospheric layers and/or a warm dusty disk/envelope provides the most likely explanations, the two of them consistent with an age in the interval 20--300 Myr. We also present new and accurate separate proper motion measurements for \object{G\,196--3\,A} and B confirming that both objects are gravitationally linked and share the same motion within a few mas\,yr$^{-1}$. After integration of the combined spectrophotometric spectral energy distributions, we obtain that the difference in the bolometric magnitudes of \object{G\,196--3\,A} and B is 6.15\,$\pm$\,0.10 mag. Kinematic consideration of the Galactic space motions of the system for distances in the interval 15--30\,pc suggests that the pair is a likely member of the Local Association, and that it lay near the past positions of young star clusters like $\alpha$ Persei less than 85~Myr ago, where the binary might have originated. At these young ages, the mass of \object{G\,196--3\,B} would be in the range 12--25\,M$_{\rm Jup}$, close to the frontier between planets and brown dwarfs.
\end{abstract}


\section{Introduction}

The characterization of brown dwarfs and planetary-mass objects of known distance, metallicity, and age can provide critical tests for evolutionary models as well as empirical references for understanding the substellar population in the field, including planets orbiting stars. Confirmed members of nearby open star clusters and substellar companions to stars and massive brown dwarfs can become good targets to establish the properties of benchmark objects. Among them, those located at the nearest distances are preferred since brown dwarfs and planets are intrinsically faint and they evolve towards lower luminosities with age (e.g., D'Antona \& Mazzitelli \cite{dantona94}; Burrows et al$.$ \cite{burrows97}; Baraffe et al$.$ \cite{baraffe98}). Tens of brown dwarfs are known to orbit stars and more massive brown dwarfs, including the discovery of a 5 Jupiter-mass planet around a brown dwarf in the young TW Hya association (Chauvin et al$.$ \cite{chauvin05}). They show L and T spectral types characterized by surface temperatures below $\sim$2200 K.

One ultra-cool substellar companion to a nearby low-mass star is \object{G\,196--3\,B}, found by direct imaging and proper motion studies at a separation of $\sim$16\arcsec~from the active M2.5 star \object{G\,196--3\,A} (Rebolo et al$.$ \cite{rebolo98}). In the discovery paper, the authors discussed that this system is young with an age in the interval 20--300 Myr. The young limit is imposed by the lack of the lithium absorption feature at 670.8 nm (Christian \& Mathioudakis \cite{christian02}) in the optical spectrum of the primary star, implying that \object{G\,196--3\,A} has efficiently depleted this element by nuclear reactions. This happens at ages $\ge$20 Myr for the mass and temperature of this particular star. The intense emission lines like H$\alpha$ and other activity properties, particularly the X-ray and UV emission of the M2.5 star, are quite similar to $\alpha$~Persei and Pleiades stars of the same temperature, suggesting that \object{G\,196--3} could have a likely age of about 100 Myr. Gizis et al$.$ \cite{gizis02} imposed a conservative upper limit of 640 Myr to the age of \object{G\,196--3\,A} based on the intensity of the most relevant chromospheric emission lines. More recently, Shkolnik et al$.$ \cite{shkolnik09} have reviewed the age of \object{G\,196--3\,A} and have set it at 25--300\,Myr.

Since its discovery and due to its relative brightness ($J$ = 14.8 mag), \object{G\,196--3\,B} has been observed spectroscopically at optical and near-infrared wavelengths by several groups. Therefore, its spectroscopic properties are well defined: \object{G\,196--3\,B} is a moderate rotator ($v$\,sin\,$i$ = 10 km\,s$^{-1}$), with no H$\alpha$ emission (Mohanty \& Basri \cite{mohanty03}). Its rotation rate is quite similar to the rotational velocity of the primary star ($v$\,sin\,$i$ = 15 km\,s$^{-1}$, Gizis et al$.$ \cite{gizis02}). The lithium feature is seen with a relatively strong intensity (a few to several Angstroms, Rebolo et al$.$ \cite{rebolo98}; Mart\'\i n et al$.$ \cite{martin99}; Kirkpatrick et al$.$ \cite{kirk01}), strongly supporting its substellar nature. In addition, \object{G\,196--3\,B} displays all spectroscopic hallmarks of a low gravity atmosphere (see next Sections), thus confirming a young age. More recently, Cruz et al$.$ \cite{cruz09} has compared the optical spectrum of \object{G\,196--3\,B} to other field dwarfs of similar temperature and spectroscopic properties, and have assigned a spectral type of L3, slightly cooler than the L1 given in Mart\'\i n et al$.$ \cite{martin99} and the L2 measured by Kirkpatrick et al$.$ \cite{kirk01}. As estimated in Rebolo et al$.$ \cite{rebolo98}, the mass of \object{G\,196--3\,B} is 25$^{+15}_{-10}$ times that of Jupiter.

In this paper we focus on the description and interpretation of the photometric properties of \object{G\,196--3\,B} from the visible to the mid-infrared wavelengths (0.6--24 $\mu$m). In addition, we present new, accurate measurements of the proper motion of each member of the pair used to determine the Galactic kinematics of the system.

\section{Observations}

\subsection{Near-infrared spectroscopy}

A low-resolution spectrum of \object{G\,196--3\,B} was obtained with the Near Infrared Camera Spectrometer (NICS) and the Amici prism, mounted at the 3.6-m Telescopio Nazionale Galileo (TNG, Observatorio del Roque de Los Muchachos, Spain) on 2002 November 27. The Amici prism and a long slit with a width of 1\farcs5 allowed us to collect spectra from 0.8 up to 2.4 $\mu$m simultaneously with a resolving power $R$\,$\sim$\,30. Total integration time was 4~min divided in eight individual exposures of 30 s each. Some clouds were present during the observations, the seeing was $\sim$1\farcs5, and \object{G\,196--3\,B} was observed at an airmass of 1.1. A small dithering over two different positions along the slit was performed to allow for proper sky subtraction. A featureless hot white dwarf (G\,191$-$B2B) was also observed with the same instrumental configuration, although at a larger airmass, to correct the target data from telluric absorption. 

\begin{figure}
\includegraphics[angle=0,scale=0.355]{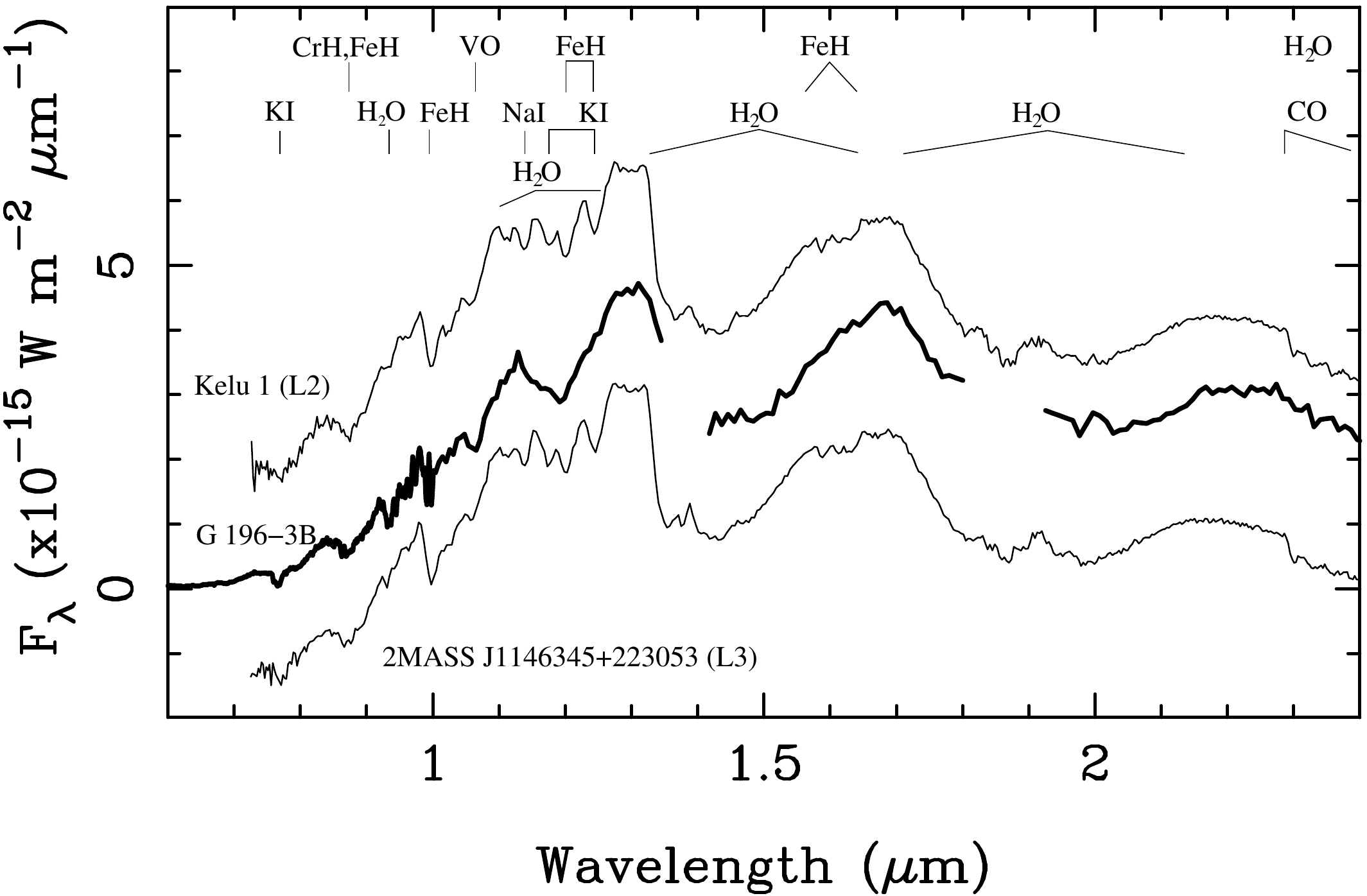}
\caption{Combined optical and near-infrared (NICS) spectrum of G\,196--3\,B (thick line) with resolving powers ($R$) of 800 (optical) and 30 (near-infrared). The vertical axis represents apparent observed fluxes of G\,196--3\,B. The SPEX prism spectra ($R \sim 120$) of the field spectroscopic standard dwarfs Kelu\,1 (Burgasser et al$.$ \cite{burgasser07}) and 2MASS\,J1146345$+$223053 (Burgasser et al$.$ \cite{burgasser10}) are also shown normalized to the peak of the $J$-band of G\,196--3\,B and shifted vertically by $\pm$1.5\,$\times$\,10$^{-15}$ flux units. The most relevant spectroscopic features are indicated. Wavelengths strongly affected by telluric absorption have been removed from G\,196--3\,B's spectrum.
\label{spectra}}
\end{figure}

The raw spectra were reduced using standard routines within the {IRAF\footnote{IRAF is distributed by the National Optical Astronomy Observatories, which are operated by the Association of Universities for Research in Astronomy, Inc., under cooperative agreement with the National Science Foundation.}} environment. NICS data were sky-subtracted, flat-fielded, aligned, combined, optimally extracted, and wavelength calibrated using the look-up table provided in the instrument's webpage\footnote{http://web.archive.org/web/20071126094610/ http://www.tng.iac.es/instruments/nics/spectroscopy.html}. We performed a fine tuning of the wavelength calibration using the deep telluric absorption features of the spectra. The instrumental response and telluric bands were removed dividing by the white dwarf observations and multiplying by the black-body spectrum of the same effective temperature (61300 K). The correction of the telluric bands was not optimal at the main telluric absorptions. NICS data were flux calibrated using integrated-light 2MASS $J$ and $H$ photometry of \object{G\,196--3\,B} and the appropriate 2MASS filter passbands. The near-infrared spectrum is combined with the optical data from Mart\'\i n et al$.$ \cite{martin99} to produce the final spectrum shown in Fig.~\ref{spectra}, which covers the wavelength interval 0.6--2.4 $\mu$m.

From the near-infrared NICS spectrum we derive spectral type L3 with an error of a subtype for \object{G\,196--3\,B} after visual inspection of our data and additional Amici spectra by Testi et al$.$ \cite{testi01} and Testi \cite{testi09}. These reference spectra were taken with the same instrument and telescope as \object{G\,196--3\,B}. We also compared our data to the SpeX-Prism spectral library maintained by A. Burgasser, confirming the L3\,$\pm$\,1 typing. Our classification basically accounts for the continuous steepening of the $J$-band slope and the increasing intensity of the H$_{2}$O absorption bands with cooler types (see also Bihain et al$.$ \cite{bihain10}). Our measurement is consistent within 1-$\sigma$ uncertainty with the typing recently assigned using optical spectra by Cruz et al$.$ \cite{cruz09}. 

The various optical and near-infrared spectroscopic features of \object{G\,196--3\,B} have been largely discussed in Mart\'\i n et al$.$ \cite{martin99}, Basri et al$.$ \cite{basri00}, Kirkpatrick et al$.$ \cite{kirk01,kirk08}, Mohanty \& Basri \cite{mohanty03}, McGovern et al$.$ \cite{mcgovern04}, McLean et al$.$ \cite{mclean07}, Allers et al$.$ \cite{allers07}, and Cruz et al$.$ \cite{cruz09}. From the NICS spectrum presented here, we confirm the narrow K\,{\sc i} lines at 1.25 $\mu$m mentioned by McLean et al$.$ \cite{mclean07}, the ``triangular'' shape of the $H$-band indicated by Allers et al$.$ \cite{allers07}, and the strong VO$+$H$_2$O absorption at around 1.2 $\mu$m mentioned by McGovern et al$.$ \cite{mcgovern04} and Allers et al$.$ \cite{allers07}. All these features altogether are indicative of youth. 

\subsection{Optical and near-infrared images}

Aimed at determining the proper motions of \object{G\,196--3\,A} and B, we collected new optical and near-infrared images on different occasions and using various telescopes: the 10-m Gran Telescopio de Canarias (GTC) and the 2.5-m Isaac Newton Telescope (INT) on Roque de los Muchachos Observatory, and the 3.5-m telescope on Calar Alto (CAHA) Observatory.  GTC data were taken on 2008 May 7 as part of the commissioning of the Acquisition and Guiding box for the telescope Nasmyth-B focus. A total of 60 frames in the Sloan $r'$ filter were obtained with the Acquisition and Slow Guiding (ASG) camera. The ASG detector was centered on the telescope optical axis to avoid field distortion and had a pixel size of 0\farcs099 projected onto the sky. Non-vignetted field of view was 1$\times$1 arcmin$^2$. Exposure time of individual images was 2 s, yielding a total on-source integration time of 120 s. Every 20 images the telescope was moved by 10\arcsec~in the north-south direction to remove the sky background contribution from the data. Observations were conducted at an average airmass of 2 and seeing of 1\farcs3. INT images were acquired using the $I$-band filter and the Wide Field Camera (WFC) on 2008 Nov 28. Each of the four CCDs of the WFC had a pixel size of 0\farcs37 and covered an area of 12$\times$24 arcmin$^2$. The seeing of the observations was 1\arcsec~and integration time was 300 s. Near-infrared $K_s$-band images were obtained with the Omega-2000 instrument (0\farcs45\,pix$^{-1}$, 15$\times$15 arcmin$^2$) attached to the 3.5-m CAHA telescope on 2010 Jan 30.  CAHA observations consisted of 6 dithers  with an offset of 10\arcsec~and on-source integration time of 31 (co-adds)$\times$1.62 s. Observing conditions were photometric and the seeing was better than 0\farcs9.

All data were reduced following standard techniques for visible and near-infrared wavelengths. Frames were flat-fielded, sky subtracted, registered against the first frame in the series, and median combined within the IRAF environment to produce the final images used in our astrometric study. Further details on INT/WFC data reduction are given in B\'ejar et al$.$ \cite{bejar99}. \object{G\,196--3\,A} is strongly saturated in the optical images, thus rendering the new optical data useful for the proper motion of the companion only. We calibrated the CAHA/Omega-2000 image astrometricaly using more than 50 sources in common with the 2MASS catalog (Skrutskie et al$.$ \cite{skrutskie06}); the rms of the plate solution was 0\farcs11. Although \object{G\,196--3\,A} slightly saturates in the CAHA/Omega-2000 data, IRAF was able to derive a precise centroid for the star. \object{G\,196--3\,B} is detected in all these images with a signal-to-noise ratio larger than 500.

\subsection{Mid-infrared photometry \label{mir}}

To measure the mid-infrared fluxes of \object{G\,196--3\,A} and B, we downloaded public images at 3.6, 4.5, 5.8, and 8.0 $\mu$m obtained with {\sl Spitzer}'s Infrared Array Camera (IRAC; Fazio et al$.$ \cite{fazio04}), and images at 24 $\mu$m obtained with the Multiband Imaging Photometer for {\sl Spitzer} (MIPS; Rieke et al$.$ \cite{rieke04}) from the {\sl Spitzer} database. We considered the observations taken for the program ID 20795, corresponding to the Astronomical Observation Request (AOR) identification numbers of 15176192 (IRAC) and 15179776 (MIPS). The IRAC images were taken on 2005 Nov 26, and the MIPS data were acquired on 2005 Nov 11. Exposure times were 26.8 s (IRAC) and 31 s (MIPS). Raw data were reduced with the {\sl Spitzer} Science Center S18.7.0 (IRAC) and S16.1.0 (MIPS) pipelines, which produced processed images with plate scales of 0\farcs6 pix$^{-1}$ (IRAC) and 2\farcs45 pix$^{-1}$ (MIPS). Processed data are given in units of MJy\,sr$^{-1}$, and for the photometric analysis we transformed them to $\mu$Jy after multiplication by factors of 8.4616 (IRAC) and 141.08 (MIPS). 

\object{G\,196--3\,A} and B are resolved in all {\sl Spitzer} images. We measured aperture photometry for each object using the task PHOT within the IRAF environment. For IRAC, we used an aperture radius of 4 pix and inner and outer radii of 24 and 40 pix for the sky annulus. For MIPS, we adopted radii of 1.22, 8.2, and 13.1 pix for the aperture and the inner and outer boundaries of the sky annulus, respectively. We applied the appropriate aperture corrections for each band to obtain the final fluxes that were converted into magnitudes using the following zero-point fluxes (taken from the IRAC and MIPS Data Handbooks): 280.9 Jy ([3.6]), 179.7 Jy ([4.5]), 115.0 Jy ([5.8]), 64.13 Jy ([8.9]), and 7.14 Jy ([24]). Our photometry is presented in Table~\ref{phot}; error bars take into account small variations in the sky contribution, and do not include the uncertainties in the calibrations of IRAC and MIPS. 

\section{Results}

\subsection{Optical and infrared colors}

\begin{figure*}
\includegraphics[angle=0,scale=0.47]{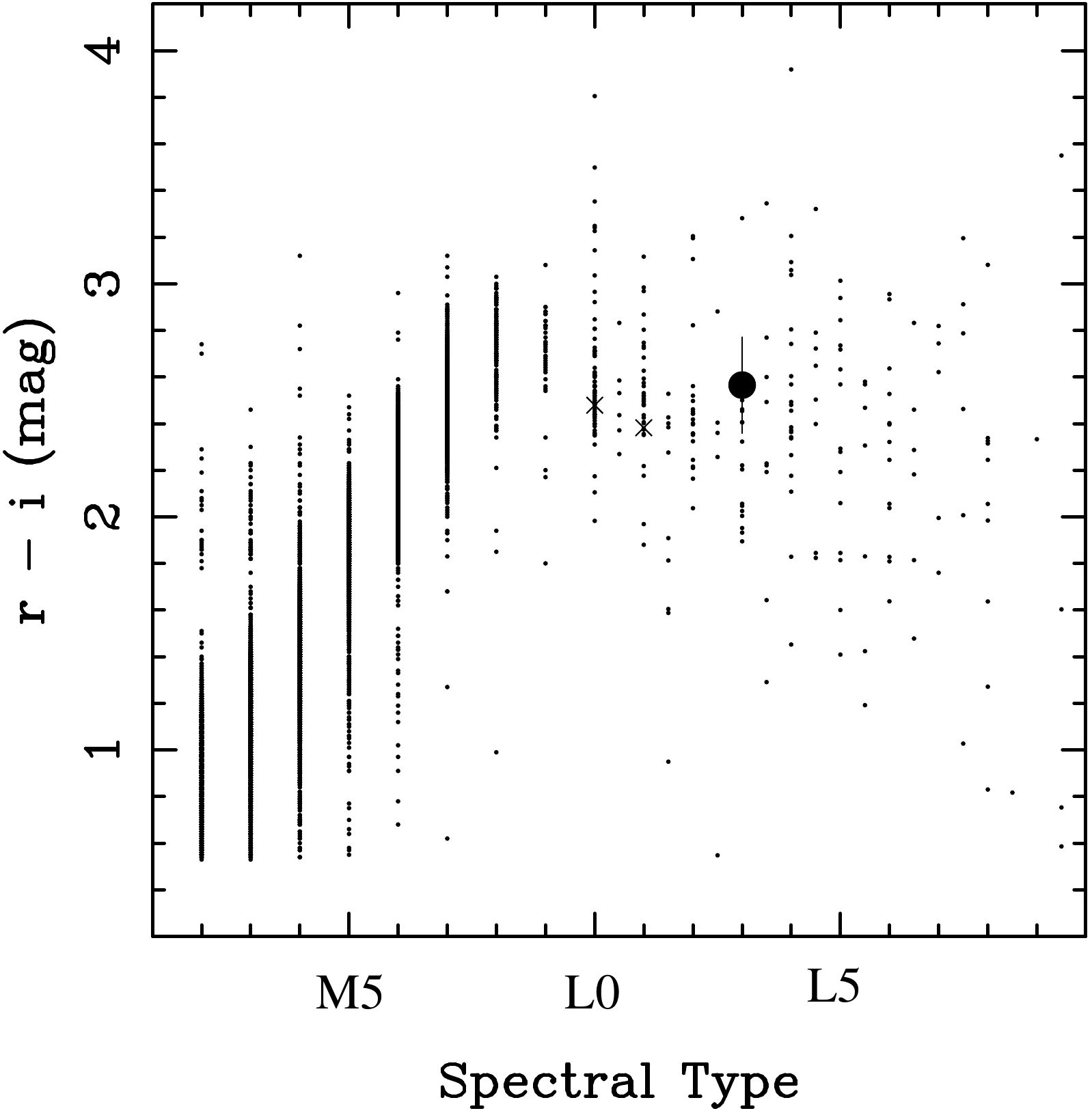}
\includegraphics[angle=0,scale=0.47]{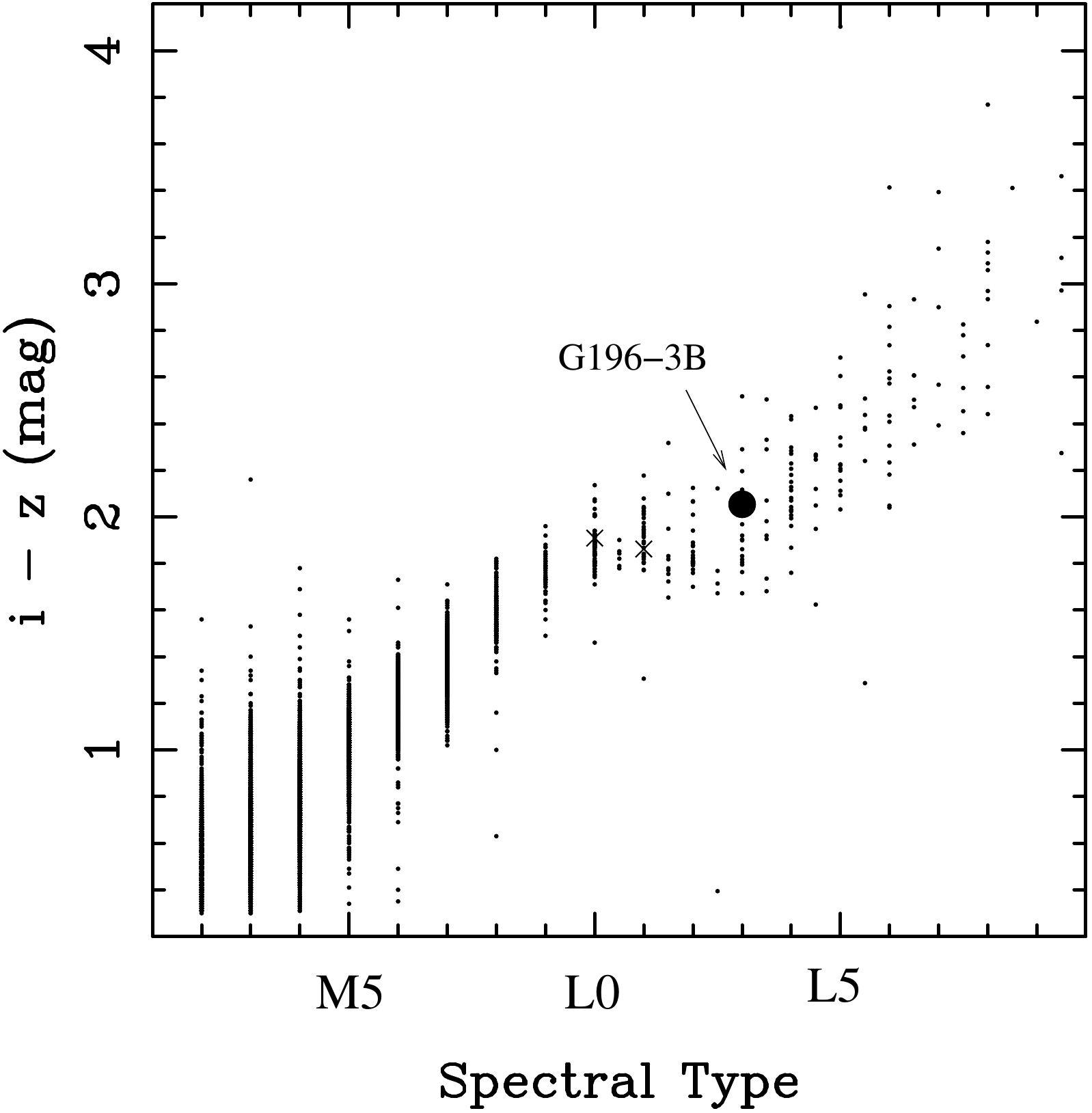}
\includegraphics[angle=0,scale=0.47]{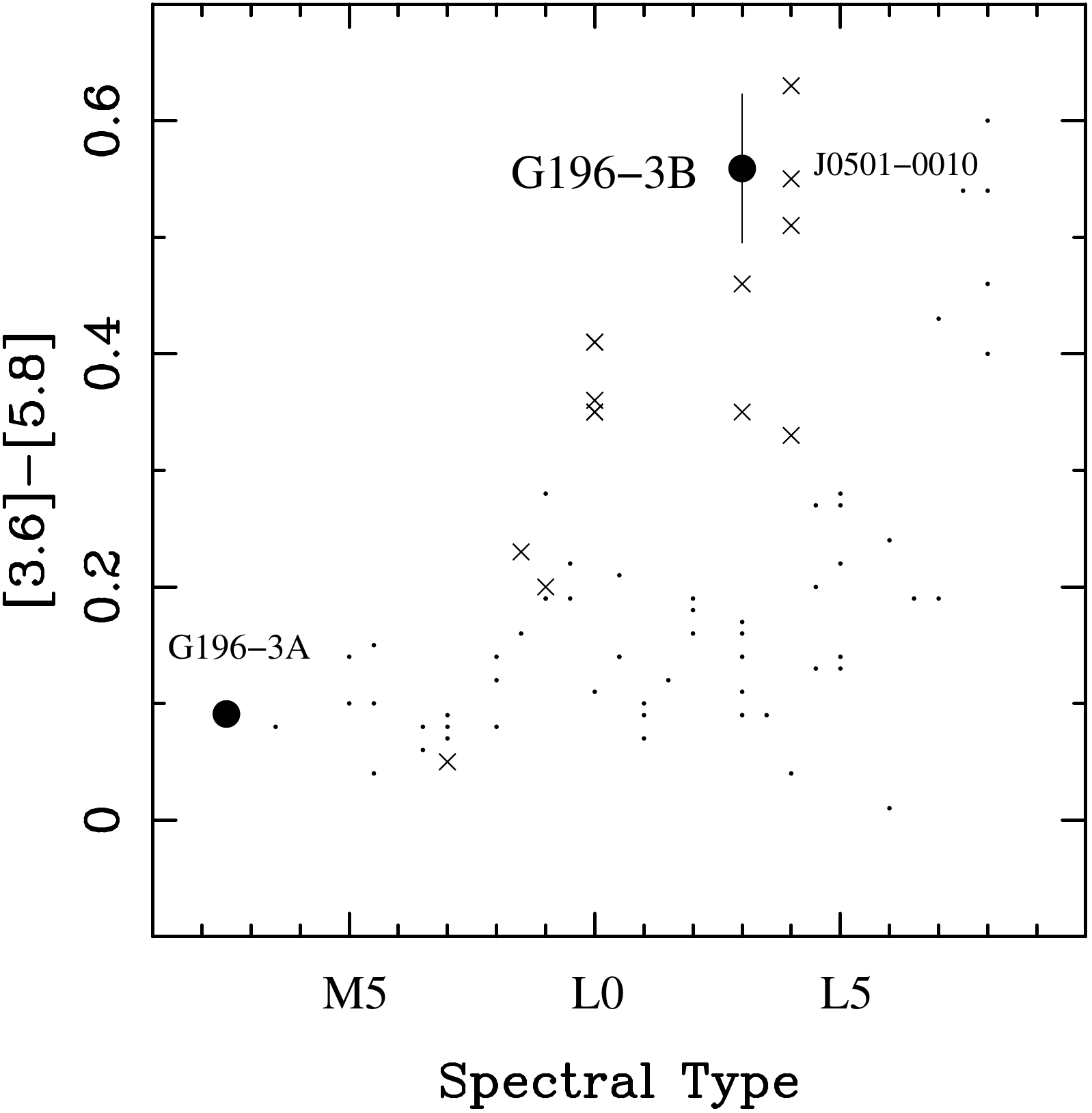}
\includegraphics[angle=0,scale=0.47]{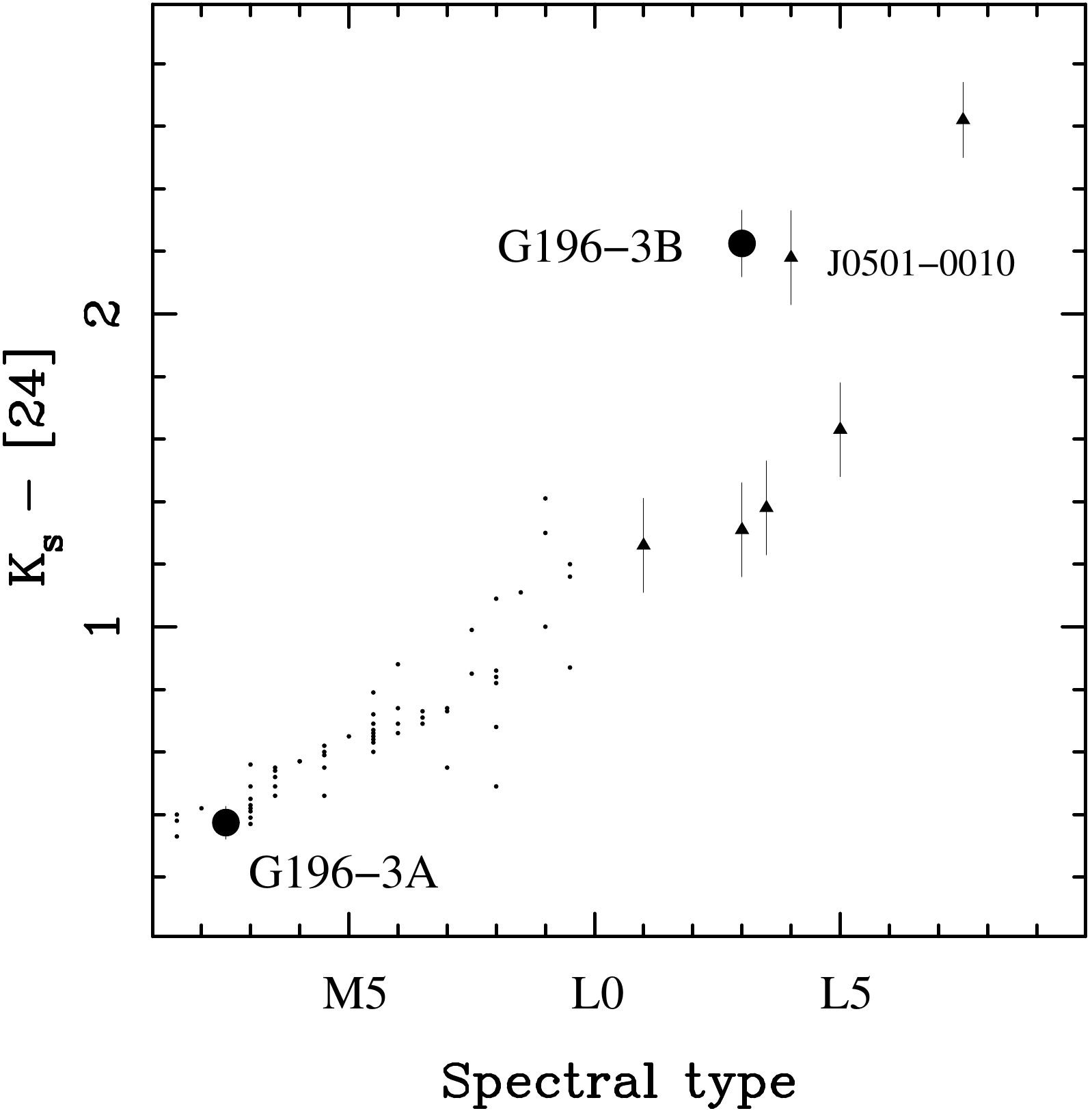}
\caption{Color versus M1--L9 spectral type. Photometry for the two components of G\,196--3 are labeled and shown with filled circles. Field dwarfs are plotted as tiny dots, crosses, and triangles. Sloan magnitudes of M and L0 dwarfs are taken from West et al$.$ \cite{west08}; Sloan data for over 250 L dwarfs were retrieved from the SDSS DR7 catalog. {\sl Spitzer}/IRAC photometry of M and L dwarfs (tiny dots in the bottom left diagram) is provided by Patten et al$.$ \cite{patten06}, while the photometry of spectroscopically confirmed young L dwarfs in the field (Luhman et al$.$ \cite{luhman09}; Allers et al$.$ \cite{allers10}) is depicted with the cross symbol. In the bottom right diagram, the {\sl Spitzer}/MIPS magnitudes of M dwarfs are taken from Gautier et al$.$ \cite{gautier07}. Our $[24]$ photometry of field L dwarfs is plotted as filled triangles. Also labeled is the young field L4-type object DENIS\,J050124.1$-$001045, common to the two panels displaying {\sl Spitzer} data.
\label{colortsp}}
\end{figure*}

Figure~\ref{colortsp} shows the Sloan, near- and mid-infrared colors of \object{G\,196--3\,A} and B as a function of spectral type. The Sloan optical photometry was retrieved from the Sloan Digital Sky Survey Data Release 7 (DR7, Abazajian et al$.$ \cite{abazajian07}). Near-infrared magnitudes are in the 2MASS photometric system (Skrutskie et al$.$ \cite{skrutskie06}), and the mid-infrared data belong to this paper. The primary star is saturated in the Sloan survey, and \object{G\,196--3\,B} is detected in the $r$, $i$, and $z$ filters with error bars smaller than 0.2 mag. We list in Table~\ref{phot} the Sloan photometry of \object{G\,196--3\,B}.

To put  \object{G\,196--3} into context, we included the photometry of field M and L dwarfs in all four panels of Fig.~\ref{colortsp}. West et al$.$ \cite{west08} provided the Sloan data for M and L0 dwarfs plotted in the upper panels, and Gautier et al$.$ \cite{gautier07} obtained the {\sl Spitzer}/MIPS photometry for field M dwarfs shown in the bottom right panel. To extend the corresponding spectrophotometric sequences into the L domain, we cross-correlated the 600 L0--L9 dwarfs known to date against Sloan DR7 (cross-correlation radius of 2\arcsec) finding over 250 positive matches, and we retrieved archived {\sl Spitzer}/MIPS images for several field L-type sources to measure their [24] brightness following the same procedure as explained in Sec.~\ref{mir}. We provide our [24] measurements and the corresponding AOR numbers in Table~\ref{mips}. Finally, the bottom left panel of Fig.~\ref{colortsp} displays the {\sl Spitzer}/IRAC sequence delineated by the M and L objects of Patten et al$.$ \cite{patten06}. 

As illustrated in Fig.~\ref{colortsp}, the optical colors of \object{G\,196--3\,B} (L3) coincide with those of the field dwarfs of similar spectral type. The $i-z$ index (top right panel of Fig~\ref{colortsp}) shows a nearly monotonic increasing trend with cooler temperature from M5 through L9, which contrasts with the flattening (or saturation) of the $r-i$ color (top left panel) at around M8. 

The near- and mid-infrared colors of \object{G\,196--3\,A} (M2.5) are also in agreement with the observations of other stars of identical classification. A different behavior is found for the substellar companion \object{G\,196--3\,B}, which systematically appears redder than the field early- to mid-L dwarfs in all near- and mid-infrared colors. From the bottom left panel of Fig.~\ref{colortsp}, \object{G\,196--3\,B} has a $[3.6]-[5.8]$ index that clearly deviates by 0.4 mag from the average trend delineated by Patten et al$.$ \cite{patten06} dwarfs (small dots), even though the field objects show a significant scatter. The red $J-K$ nature of young field L dwarfs is also noted by Cruz et al$.$ \cite{cruz09}. These authors reported on the optical spectroscopic features that are a signpost of low-gravity atmospheres for all young L-type dwarfs in the field, including \object{G\,196--3\,B}. Luhman et al$.$ \cite{luhman09} provided the IRAC photometry for some of these young L dwarfs; in addition, Allers et al$.$ \cite{allers10} reported the discovery of a young L dwarf binary in the field. We have included Luhman et al.'s and Allers et al.'s data (cross symbols) in various panels of Fig.~\ref{colortsp}. One young field L4-type source, DENIS\,J050124.1$-$001045 (Reid et al$.$ \cite{reid08}), is in our sample of new [24] measurements. The location of \object{G\,196--3\,B} in the color-spectral type diagrams of Fig.~\ref{colortsp} is not special when compared to the ``family'' of field, spectroscopically-confirmed young L dwarfs. We note, however, that its colors lie at the reddest edge of the color--spectral type distributions. The near- and mid-infrared indices of \object{G\,196--3\,B} resemble those of the field dwarfs with the latest L-types, in contrast to the early-L classification obtained from optical and near-infrared spectra. 

\subsection{Spectral energy distribution}

\begin{figure}
\includegraphics[angle=0,scale=0.47]{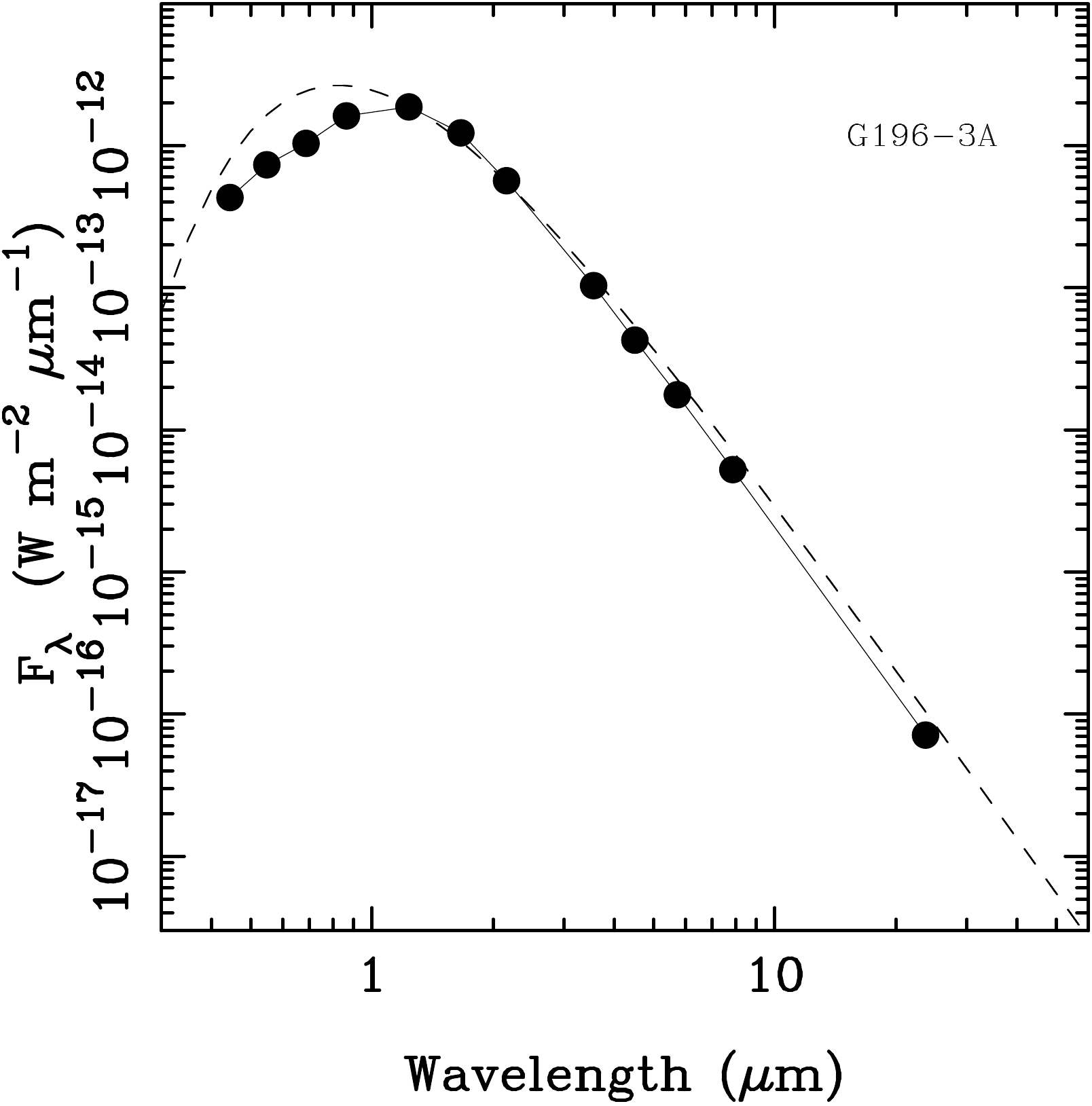}
\includegraphics[angle=0,scale=0.47]{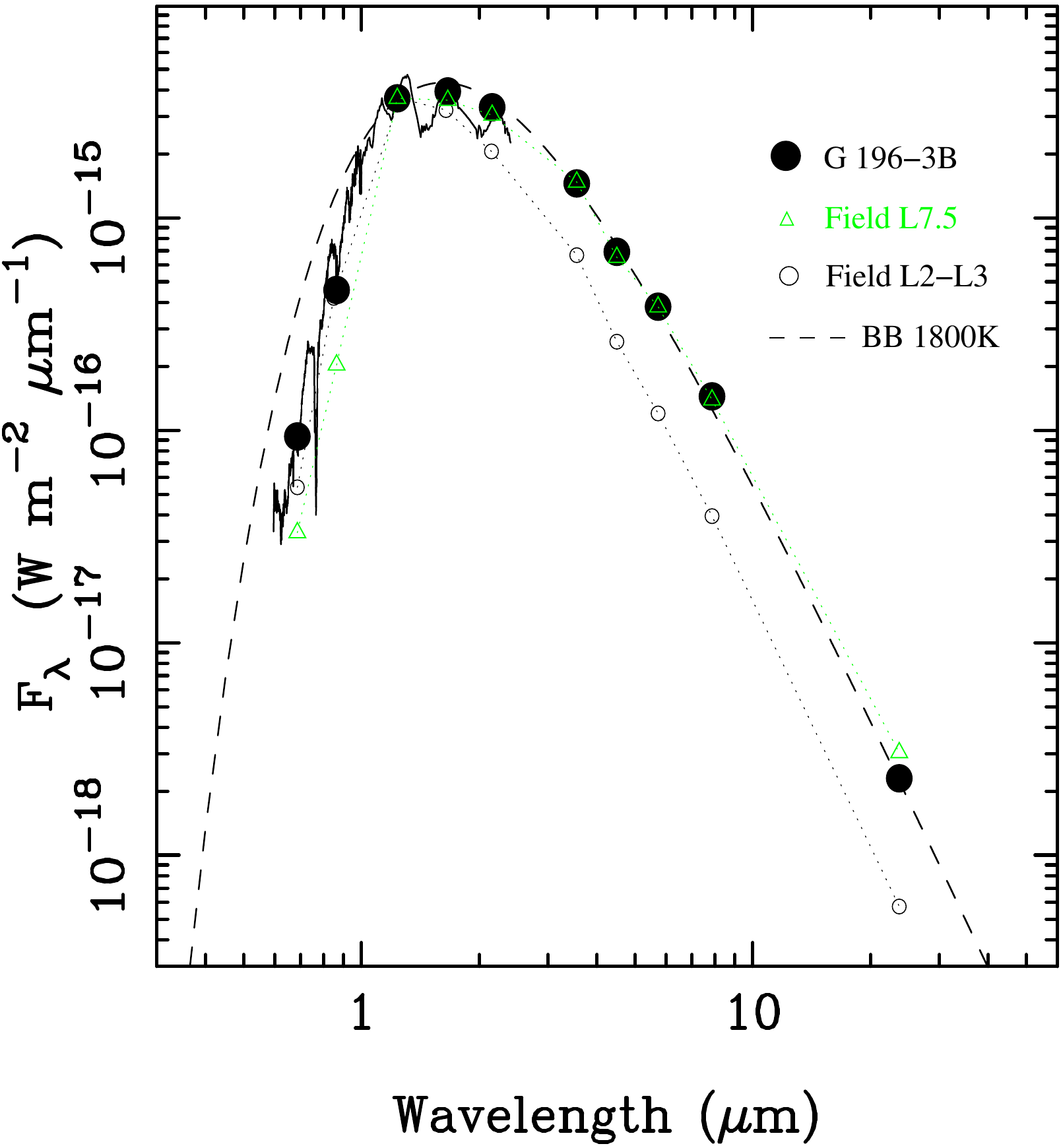}
\caption{Photometric spectral energy distributions (SEDs, filled dots) of G\,196--3\,A {\sl (top panel)} and B {\sl (bottom panel)}. Also in the bottom panel, the optical and near-infrared spectra of G\,196--3\,B (solid line) and the average photometric SEDs of field L2--L3 dwarfs (open circles) and 2MASS\,J0825196$+$211552 (L7.5, triangles) are plotted. Dashed lines correspond to the emission of black bodies of temperatures 3500\,K {\sl (top panel)} and 1800\,K {\sl (bottom panel)}. All SEDs are normalized to the $J$-band fluxes of G\,196--3\,A and B.\label{sed}}
\end{figure}

We combined all available broad-band photometry, $B\,V\,R\,I\,J\,H\,K_s\,[3.6]\,[4.5]\,[5.8]\,[8.0]\,[24]$, to create 0.44--24.0 $\mu$m (\object{G\,196--3\,A}) and 0.68--24\,$\mu$m (\object{G\,196--3\,B}) photometric spectral energy distributions (SEDs). For \object{G\,196--3\,B} we also have the combined optical and near-infrared spectrum shown in Fig.~\ref{spectra}, which covers the wavelength range 0.6--2.4 $\mu$m. Observed magnitudes were converted to monochromatic fluxes at the central wavelength of each filter using the zero point fluxes provided in the literature for 2MASS and those mentioned in Sec.~\ref{mir} for {\sl Spitzer} instruments. The computed SEDs normalized to the $J$-band are illustrated in Fig.~\ref{sed}. For comparison with \object{G\,196--3\,B}, we overplotted the SED of the L7.5 dwarf 2MASS\,J0825196$+$211552 (Kirkpatrick et al$.$ \cite{kirk00}) and the average photometric SED of L2--L3 dwarfs. These were obtained in a similar manner to \object{G\,196--3}. We compiled the required photometry $R$ through $[8.0]$ from the 2MASS catalog, Liebert \& Gizis \cite{liebert06}, and Patten et al$.$ \cite{patten06}. The [24] data come from this paper. All these data were collected with the same instruments and telescopes (2MASS and {\sl Spitzer}/IRAC/MIPS) as for the pair \object{G\,196--3}, except for the optical $R\,I$ photometry (Rebolo et al$.$ \cite{rebolo98}). \object{G\,196--3\,B} shows an $I-J$ color quite similar to that of the field L2--L3 dwarfs. However, its near- and mid-infrared SED has a shape more similar to that of the much cooler field L7.5 dwarf.  

The SED of \object{G\,196--3\,A} clearly indicates its photospheric origin since there is no obvious mid-infrared flux excesses up to 24\,$\mu$m. To detect a debris disk if it exists surrounding this low-mass star, observations at wavelengths longer than 24\,$\mu$m should be undertaken. The SED of \object{G\,196--3\,B} displays a quite smooth pattern with no clear evidence for a rising flux emission or changing slope at the reddest wavelengths typical of disks significantly cooler than the central object. However, this SED appears overluminous longwards of the $H$-band when compared to the average SEDs of field L2--L3 dwarfs (Fig.~\ref{colortsp}). We discuss different possible explanations in Sec.~\ref{discussion}.

\subsection{Bolometric Magnitudes}

We integrated the observed SEDs over wavelength to obtain the apparent bolometric magnitudes ($m_{\rm bol}$)$.$ To estimate the flux contribution at wavelengths bluer and redder than the available photometry, we fitted a black-body spectrum to the SEDs of \object{G\,196--3\,A} and B, obtaining that spherical black bodies with temperatures of 3500 and 1800~K provide reasonable match to the observed fluxes in the red. The black-body emissions normalized at the $J$-band are shown in Fig.~\ref{sed}. We integrated the extended SEDs over the wavelength interval 0.1--1000 $\mu$m using the simple trapezoidal rule, and applied $m_{\rm bol}$\,=\,$-2.5$\,log\,$f_{\rm bol}$ $-$\,18.988 (Cushing et al$.$ \cite{cushing05}, where $f_{\rm bol}$ is in units of W\,m$^{-2}$) to obtain the following apparent bolometric magnitudes: $m_{\rm bol}$\,=\,9.85\,$\pm$\,0.10 mag for \object{G\,196--3\,A}, and $m_{\rm bol}$\,=\,16.00\,$\pm$\,0.10 mag for \object{G\,196--3\,B} (Table~\ref{phot}). The uncertainties take into account the photometric error bars, the errors in the zero point fluxes, and an uncertainty of $\pm$100~K in the temperature of the black bodies used to extrapolate the SEDs. The bolometric magnitude difference between \object{G\,196--3\,B} and A turns out to be 6.15\,$\pm$\,0.10 mag.

Our measured $m_{\rm bol}$ of \object{G\,196--3\,A} is in excellent agreement with $m_{\rm bol}$\,=\,9.82\,$\pm$\,0.07 mag determined using the various bolometric corrections (BCs) available in the literature for field stars with the spectral classification M2.5V (e.g., Pickles \cite{pickles85}; Kirkpatrick et al$.$ \cite{kirk93}; Leggett et al$.$ \cite{leggett00}). This implies that the TiO, CO, H$_2$O, and other molecular absorptions present in the star's spectrum have weak effect on our $m_{\rm bol}$ determination. 

The $m_{\rm bol}$ of \object{G\,196--3\,B} derived using BC$_J$ of field L2--L3 dwarfs, $m_{\rm bol}$\,=\,16.64\,$\pm$\,0.15 mag (Reid et al$.$ \cite{reid01}; Dahn et al$.$ \cite{dahn02}; Vrba et al$.$ \cite{vrba04}) is different to the one $m_{\rm bol}$\,=\,16.09\,$\pm$\,0.10 mag obtained from BC$_K$ (Leggett et al$.$ \cite{leggett01}; Golimowski et al$.$ \cite{golimowski04}; Vrba et al$.$ \cite{vrba04}). This is likely due to the very red $J-K_s$ color of \object{G\,196--3\,B}, which significantly deviates from that of L2--L3 dwarfs. Both determinations are fainter than our measurement by $\sim$0.6 ($J$) and $\sim$0.1 mag ($K$). Part of this discrepancy might also arise from our method to determine bolometric magnitudes, e.g., the use of black bodies instead of synthetic spectra to complete the observed SED below 0.6 $\mu$m and above 24 $\mu$m, or the presence of strong absorption bands in the spectrum. None of these appears to be relevant for the M2.5 star. Leggett et al$.$ \cite{leggett01} used a similar approach to ours for bolometric magnitude determination and also employed synthetic model atmospheres for completing the SEDs, finding that no correction is needed for late-M to mid-L dwarfs. To test this, we integrated the pure photometric SED and the combined photometric and spectroscopic SED of \object{G\,196--3\,B} finding good agreement in the bolometric magnitudes (15.91 and 16.00 mag, respectively) within the $\pm$0.10 dex error bars. We adopt our calculated $m_{\rm bol}$ for \object{G\,196--3\,B} (Table~\ref{phot}), from which we infer the following bolometric corrections: BC$_J$\,=\,1.16\,$\pm$0.10 and BC$_K$\,=\,3.22\,$\pm$\,0.10 mag.

\subsection{Proper motion}

We derived accurate proper motions independently for \object{G\,196--3\,A} and B using a large battery of images and astrometric databases available to us, including our own published and new data. For the primary star, there are a total of 12 astrometric epochs (Table~\ref{mpA}) covering observations over 111.8 yr. Our measurement, $\mu_\alpha\,{\rm cos}\delta$ = $-$142.3\,$\pm$\,0.6 mas\,yr$^{-1}$, $\mu_\delta$ = $-$197.0\,$\pm$\,0.9 mas\,yr$^{-1}$, derived as in Caballero \cite{caballero10}, coincides within 1-$\sigma$ uncertainty with the astrometric values provided by, e.g., Salim \& Gould \cite{salim03}, L\'epine \& Shara \cite{lepine05}, and R\"oser et al$.$ \cite{roser08}. The quoted error bars, obtained as the standard deviation of the linear fit to the $\alpha$ and $\delta$ coordinates, are more than a factor of 2.5 smaller than those of previous measurements.

There are less astrometric epochs spanning 12 yr of observations between 1998 and 2010 that can be used to derive a reliable proper motion for \object{G\,196--3\,B}. Besides 2MASS and Sloan images (Table~\ref{mpA}), where the companion is clearly detected, Table~\ref{mpB} provides additional images we employed in our astrometric analysis. By comparing the positions of \object{G\,196--3\,B} relative to various reference stars in the field and to the NOT data, we determined $\mu_\alpha\,{\rm cos}\delta$ = $-$144\,$\pm$\,2 mas\,yr$^{-1}$ and $\mu_\delta$ = $-$190\,$\pm$\,20 mas\,yr$^{-1}$ for \object{G\,196--3\,B}. The uncertainties represent the standard deviation of individual measurements. The agreement between our proper motion and that derived by Jameson et al$.$ \cite{jameson08} is better than 1-$\sigma$. Both \object{G\,196--3\,A} and B share the same proper motion within a few mas\,yr$^{-1}$, thus providing strong support for the true gravitational companionship of the pair. 

The accuracy of our astrometry did not allow us to detect any orbital displacement in the 12-yr interval between 1998 Feb (NOT/ALFOSC) and 2010 Jan (CAHA/Omega-2000). We measured the projected angular separation of the binary at 15\farcs99\,$\pm$\,0\farcs06, and the position angle at 209.2\,$\pm$\,0.3 deg. Error bars represent the standard deviation of the various individual measurements. Systematic errors can be larger than those quoted here.

\section{Discussion \label{discussion}}

\subsection{Hertzsprung-Russell diagram}

To construct the Hertzsprung-Russell (HR) diagram we need to convert observed $m_{\rm bol}$ and spectral type into luminosity and $T_{\rm eff}$. No parallactic distance is known for \object{G\,196--3}; however, we can set lower and upper ``spectroscopic'' limits by assuming different ages and binary status for each component. The $m_{\rm bol}$ of \object{G\,196--3\,A} and B will be compared to the absolute magnitudes of objects of similar spectral type and well-determined parallax, including field sources and members of stellar clusters. Our reference star clusters are the 120-Myr Pleiades (catalogs by Stauffer et al$.$ \cite{stauffer07} and Bihain et al$.$ \cite{bihain10}), the 50-Myr IC\,2391 (catalog by Barrado y Navascu\'es et al$.$ \cite{barrado01}), and the 3-Myr $\sigma$\,Orionis (catalogs by Zapatero Osorio et al$.$ \cite{osorio00}, \cite{osorio02}, and Sacco et al$.$ \cite{sacco08}). Table~\ref{dist} provides our distance estimates with an uncertainty of 10\%~for two possible scenarios: single and equal-mass objects. The minimum and maximum possible distances to \object{G\,196--3} are 15 and 51\,pc for ages typical of the field and 3\,Myr. From considerations largely discussed in Rebolo et al$.$ \cite{rebolo98} and Shkolnik et al$.$ \cite{shkolnik09}, the age of \object{G\,196--3} is possibly in the range 20--300 Myr, implying a likely distance in the interval 18--30 pc for the single object scenario and 25--40 for the binary one. The projected physical separation of the pair thus turns out to be 285--640 AU.

Rebolo et al$.$ \cite{rebolo98} estimated $T_{\rm eff}$ values for both \object{G\,196--3\,A} and B. There are new $T_{\rm eff}$--spectral type relations available in the literature, particularly for the L dwarfs (Luhman \& Rieke \cite{luhman98}; Allen \cite{allen00}; Leggett et al$.$ \cite{leggett01}, \cite{leggett02}; Dahn et al$.$ \cite{dahn02}; Golimowski et al$.$ \cite{golimowski04}; Vrba et al$.$ \cite{vrba04}; Cushing et al$.$ \cite{cushing08}; \cite{stephens09}). We used them to update our measurements, obtaining 3480\,$\pm$\,95~K for the M2.5 star and 1870\,$\pm$\,100~K for the L3 substellar companion (Table~\ref{phot}). The error bars account for the dispersion of the various $T_{\rm eff}$--spectral type relations and an uncertainty of a subclass in the spectral type classification. These measurements are consistent, although slightly warmer, within 1-$\sigma$ uncertainty with those of Rebolo et al$.$ \cite{rebolo98}. Basri et al$.$ \cite{basri00} derived a higher $T_{\rm eff}$ (2200--2400 K) for \object{G\,196--3\,B} based on the spectroscopic fitting of the Cs\,{\sc i} and Rb\,{\sc i} atomic lines with ``high-gravity'' synthetic model atmospheres. At this warm temperature, one would expect the presence of TiO bands in optical spectra. However, the spectra of \object{G\,196--3\,B} shown in Rebolo et al$.$ \cite{rebolo98}, Mart\'\i n et al$.$ \cite{martin99}, and Cruz et al$.$ \cite{cruz09} lack TiO bands, and only faint VO, CrH, FeH molecular absorption remains. This suggests a cooler temperature for {G\,196--3\,B}, a fact already acknowledged by Basri et al$.$ \cite{basri00} and in agreement with our estimate. 

\begin{figure}
\includegraphics[angle=0,scale=0.47]{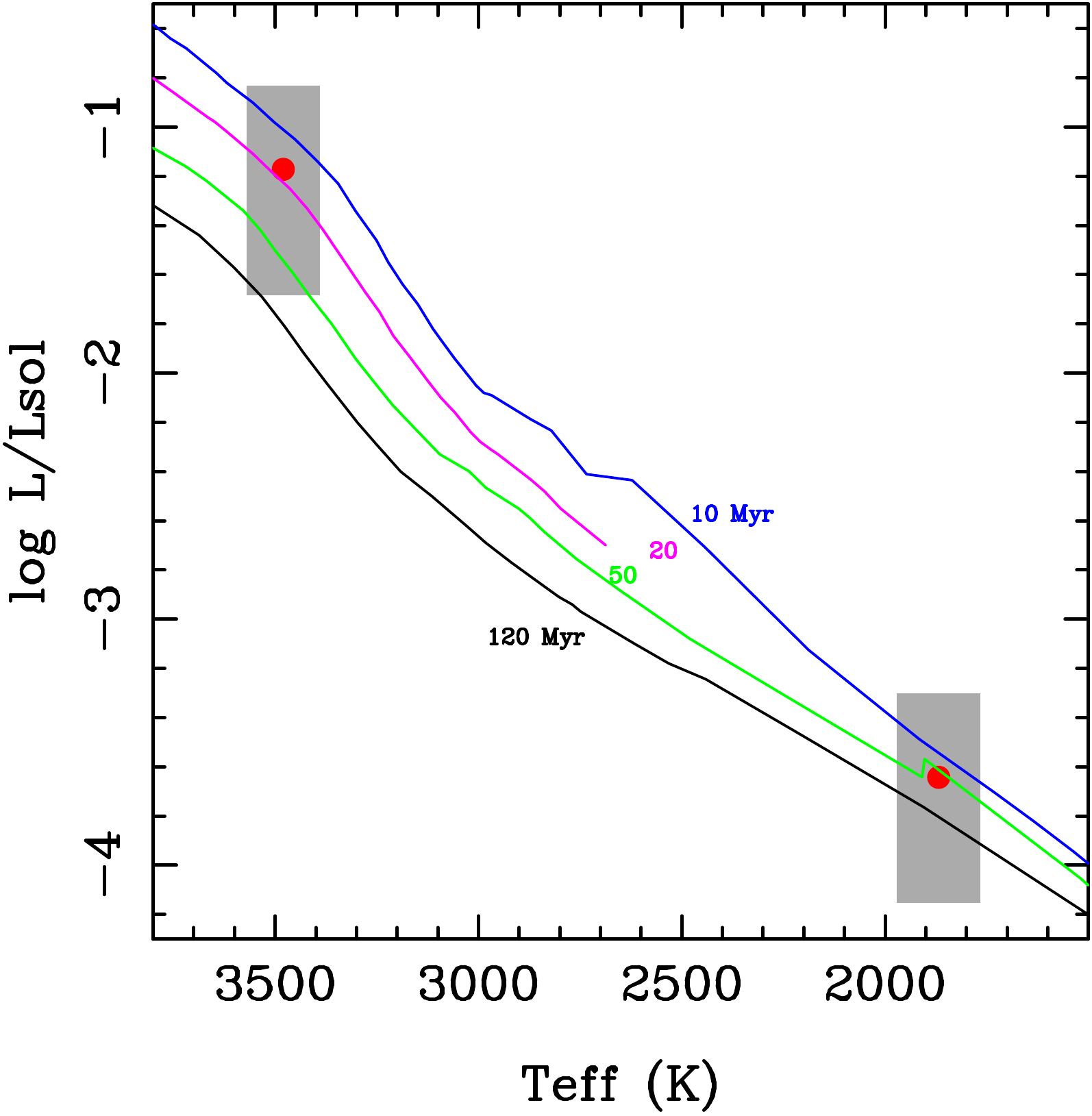}
\caption{Hertzsprung-Russell diagram for G\,196--3. The grey-shaded regions indicate the expected locations of both components for a distance interval 15--40\,pc. The position of the pair at the (arbitrary) distance of 27\,pc is plotted as red filled circles. Overplotted are theoretical isochrones of 10, 20, 50, and 120 Myr (Baraffe et al$.$ \cite{baraffe98}; Chabrier et al$.$ \cite{chabrier00}). The small ``bump'' observed at around 1800--1900 K in the 50-Myr isochrone is due to deuterium nuclear burning. 
\label{hr}}
\end{figure}

Luminosities (for the distance interval 15--40~pc, i.e., single objects) and $T_{\rm eff}$s are used to place \object{G\,196--3\,A} and B in the HR diagram displayed in Fig.~\ref{hr}. Overplotted are theoretical isochrones (10, 20, 50, and 120 Myr) produced by the Lyon group (Baraffe et al$.$ \cite{baraffe98}; Chabrier et al$.$ \cite{chabrier00}). Because \object{G\,196--3\,A} and B are supposed to be coeval, one isochrone should reproduce the relative location of both objects if they are single. In Fig.~\ref{hr}, we also indicate the positions of \object{G\,196--3\,A} and B at one given distance; \object{G\,196--3\,A} appears overluminous with respect to the isochrone describing the companion, which may suggest that this star is a close binary, or \object{G\,196--3\,B} appears underlminous compared to the isochrone passing through the primary star. Without a definitive, accurate astrometric parallax, the detailed study of the HR diagram of the pair is not possible. For the age interval 20--300 Myr, the most likely masses of the two components are 0.40--0.55 $M_\odot$ and 0.012--0.040$M_\odot$ for \object{G\,196--3\,A} and B. If the age of the system is around 20--80\,Myr (see next Section), \object{G\,196--3\,B} would have a mass at around the planet--brown dwarf frontier, and it would be burning deuterium as suggested in Fig.~\ref{hr}.

\subsection{Moving groups}

\begin{figure*}
\includegraphics[angle=0,scale=0.9]{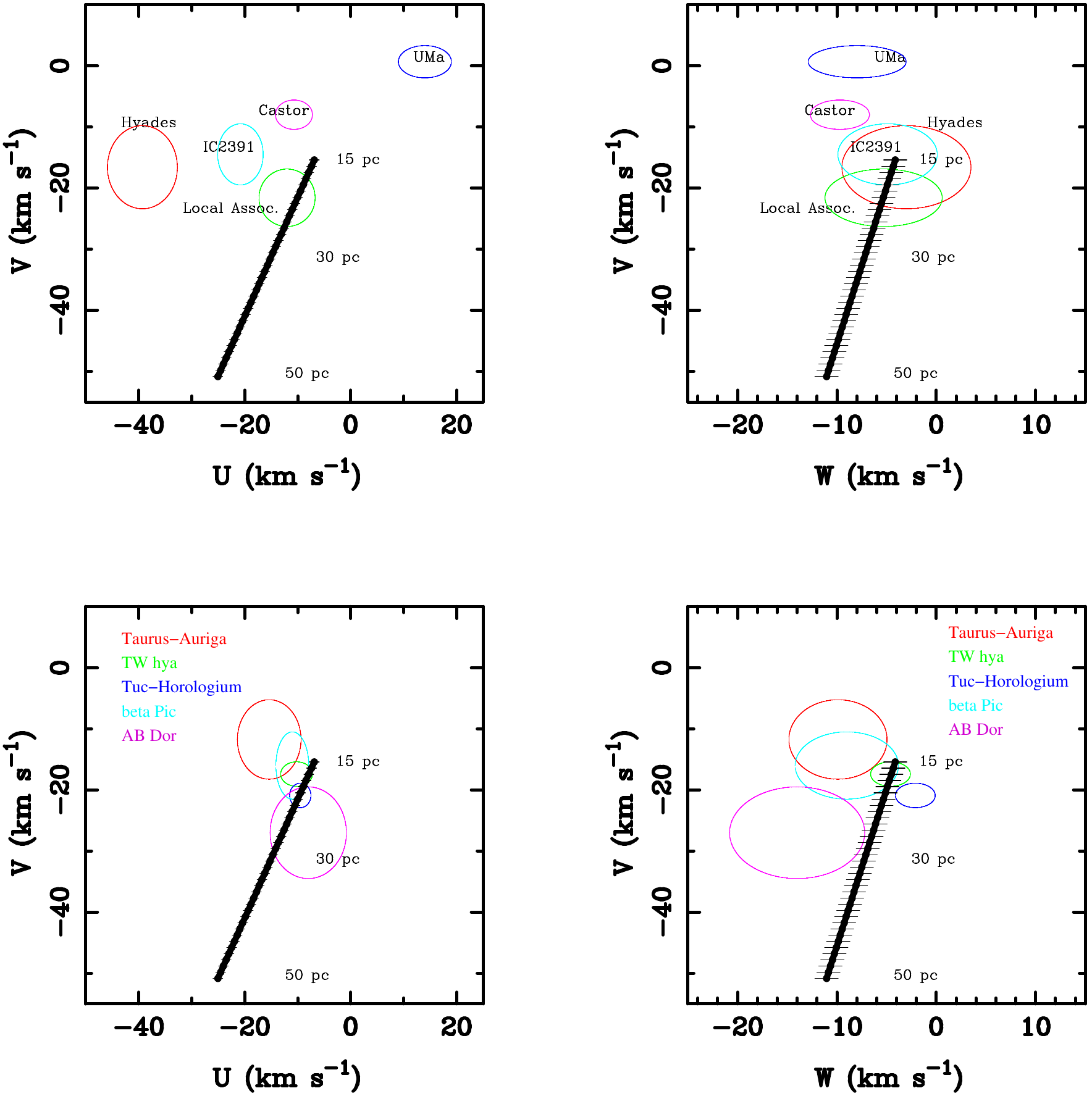}
\caption{Possible space velocities of G\,196--3 (black straight line) for the distance interval 15--50 pc. Overplotted are the ellipsoids of known young star associations and moving groups. Galactocentric $U$ velocity is positive towards the Galactic center.
\label{uvw}}
\end{figure*}

Young objects of less than a few hundred Myr can be identified as members of star-forming regions and young stellar associations (Barrado y Navascu\'es \cite{barrado98}; Montes et al$.$ \cite{montes01}; Zuckerman \& Song \cite{zuckerman04}; Torres et al$.$ \cite{torres08}). The Galactic space velocities $UVW$, X-ray luminosities, and a detailed analysis of the spectra provide reliable indicators of membership to an association. To derive $UVW$ for \object{G\,196--3\,B} we used the proper motion of the primary star, which has a very small error bar ($\le$1 mas\,yr$^{-1}$), and the radial velocity ($-1.5\,\pm\,1.0$ km\,s$^{-1}$) obtained for the substellar object by Basri et al$.$ \cite{basri00}, and applied the equations by Johnson \& Soderblom \cite{johnson87} to derive the velocities displayed in Fig.~\ref{uvw} for the distance interval 15--50 pc. The distance-dependent (therefore, age-dependent) values of $U$, $V$, and $W$ follow straight lines in the $UV$ and $VW$ planes. The uncertainties associated to all three Galactic velocities for a given distance come from the proper motion and radial velocity error bars. 

Figure~\ref{uvw} also shows the ellipsoids corresponding to well known young stellar moving groups of the solar neighborhood (data from Zuckerman \& Song \cite{zuckerman04}; Torres et al$.$ \cite{torres08}). The Ursa Majoris, Hyades, Castor, Taurus-Auriga, and IC\,2391 moving groups lie more than 10 km\,s$^{-1}$ away from the likely positions of \object{G\,196--3}. The Tucana-Horologium and AB Doradus moving groups are only marginally consistent with \object{G\,196--3}. Therefore, membership of the pair to any of these associations is unlikely. 

For a distance interval between 18 and $\sim$30 pc, the space motion of \object{G\,196--3} matches the space motions of the Local Association (or Pleiades moving group), TW~Hydrae, and $\beta$~Pictoris. Both TW~Hydrae and $\beta$~Pictoris have very young ages ($\sim$8--10 and 12 Myr, respectively), which are smaller than our upper limit on the age of the pair. Furthermore, TW~Hydrae is dominated by M-type stars, all of which have preserved lithium (Song et al$.$ \cite{song03}), in contrast to the spectroscopic observations of \object{G\,196--3\,A}. The only remaining moving group that could explain the $UVW$ values of \object{G\,196--3} is the Local Association, where smaller structures of different ages (all below 300 Myr) like the Pleiades (120 Myr), $\alpha$~Persei (50--80 Myr), and IC\,2602 ($\sim$70 Myr) open clusters are contained. 

We have linearly back-traced in time the Galactic orbit of \object{G\,196--3} and of 120 open star clusters from the lists by Kharchenko et al$.$ \cite{kharchenko05} and Makarov et al$.$ \cite{makarov07}. These authors provided current 3D positions (or sufficient data to properly calculate them), space motions, and ages for all clusters. We selected those with ages in the interval 1--300 Myr and distances up to several hundred parsecs, for which our linear approximation does not deviate much from true calculations including several rotations around the Galactic center. \object{G\,196--3} may have been located within a few to several ten parsec distance from the $\alpha$~Persei or Collinder~65 open clusters about $\sim$50--85 or $\sim$20--30 Myr ago, respectively. This is indeed in agreement with the most likely age of the system, suggesting that \object{G\,196--3} might have shared the same birth molecular cloud than any of these two star clusters, being ejected soon after formation. This indicates that the age of \object{G\,196--3} may be younger than the Pleiades star cluster, with a likely value between 20 and 85\,Myr. We note, however, that the age and distance parameters of Collinder~65 are not well constrained, thus introducing some uncertainty in our estimations, and that precise radial velocity and parallax of \object{G\,196--3} are necessary to refine our analysis.

\subsection{Disk/envelope or enshrouded atmosphere?}

The most striking observable property of the overall SED of \object{G\,196--3\,B} is its apparent overluminosity at near- and mid-infrared wavelengths (see Fig.~\ref{sed}), which translates into colors redder than expected for its optical and near-infrared spectral type. We propose the following explanations, one or various of which may account for this signature: different metallicity, low-gravity atmosphere, a dust-enshrouded atmosphere, continuum emission from a ``hot'' disk or dusty envelope, and the presence of a cool, unresolved companion.

Theory of model atmospheres predicts that metallicity and gravity impacts the colors of ultra-cool dwarfs (e.g., Burrows et al$.$ \cite{burrows06}; Leggett et al$.$ \cite{leggett07}). Furthermore, Knapp et al$.$ \cite{knapp04} indicated that gravity is a ``required'' parameter to understand the sequences of L and T dwarfs in color-magnitude diagrams. Reducing the metallicity increases the importance of molecular hydrogen collision-induced absorption in the infrared, producing blue colors. This is clearly observed in subdwarfs with L spectral classification (e.g., Burgasser et al$.$ \cite{burgasser03}; Cushing et al$.$ \cite{cushing09}). \object{G\,196--3\,B} shows an opposing behavior, suggesting that it might have a metal-rich atmosphere. These atmospheres are also dustier. Unfortunately, the primary star \object{G\,196--3\,A} does not contribute to the understanding of the metallic content of the pair, since for M dwarfs any reliable abundance determination is not an easy task (e.g., Bonfils et al$.$ \cite{bonfils05}). Nevertheless, the optical spectrum of \object{G\,196--3\,A} is very similar to standard M2--M3 stars, including the solar-metallicity Pleiades and $\alpha$ Persei stellar members (see Rebolo et al$.$ \cite{rebolo98}). Furthermore, the great majority of nearby stars have nearly solar composition (Nordstr\"om et al$.$ \cite{nordstrom04}; Valenti \& Fischer \cite{valenti05}). The probable membership of \object{G\,196--3} in the Local Association also adds support to the solar abundance of the system. From theoretical model atmospheres, metal-rich ultracool dwarfs have redder 1--4 $\mu$m colors than solar-metallicity dwarfs while the 5--15 $\mu$m spectrum does not change significantly (Stephens et al$.$ \cite{stephens09}), which contrasts with the properties of \object{G\,196--3B}. Therefore, high metallicity cannot explain the infrared overluminosity of \object{G\,196--3B}.

In ultra-cool dwarfs, decreasing the atmospheric gravity or gas density has a qualitatively similar effect on the photospheric gas pressure and chemistry of gas and condensates as increasing the metallicity (Lodders \& Fegley \cite{lodders02}). State-of-the-art model atmospheres indicate that low gravities (log\,$g$\,$\sim 4$ dex, cgs units) tend to produce redder infrared colors than high gravities (log\,$g$\,$\sim$\,5 dex, e.g., $J-K$, $H-K$, $K-L'$, $K-[3.5]$, and $[4.5]-[5.8]$, see Burrows et al$.$ \cite{burrows06}; Leggett et al$.$ \cite{leggett07}, \cite{leggett10}). The impact of gravity on these colors is significant (at the $\ge$20\%~level) particularly for the T dwarfs, while the colors of earlier dwarfs do not appear to be as much impacted theoretically. The exact amounts of the color deviations observed in the L3-type \object{G\,196--3\,B} are not quantitatively explained by the models. \object{G\,196--3\,B} displays all the spectroscopic hallmarks typically used to recognize low-gravity objects (see previous Sections). Furthermore, from kinematic considerations its age is likely younger than the Pleiades, thus supporting a low-gravity atmosphere. Evolutionary models predict a surface gravity in the range log\,$g$\,=\,4.1--4.7 dex for \object{G\,196--3\,B}, which compares to the higher values (log\,$g$ $\sim$\,5 dex) of older dwarfs of similar types. However, early-L dwarfs members of very young clusters like Upper Scorpius ($\sim$5 Myr; Lodieu et al$.$ \cite{lodieu08}) and $\sigma$~Orionis ($\sim$3 Myr, Zapatero Osorio et al$.$ \cite{osorio07}; Caballero et al$.$ \cite{caballero07}), for which evolutionary models predict even lower surface gravities (log\,$g$ $\le$\,4 dex), do not show colors differing significantly from the ``high-gravity'' field. It appears that the behavior of colors with gravity is not necessarily monotonic. Interestingly, the L-type substellar object 2MASS J1207334$-$393254b, which is a companion to a brown dwarf of the TW Hydrae Association (age $\sim$10 Myr, Chauvin et al$.$ \cite{chauvin05}), also displays very red $H-K$ and $K-L$ colors as compared to field dwarfs of similar types (Mohanty et al$.$ \cite{mohanty07}), suggesting that the change from not very red colors at very young ages to very red colors at intermediate ages (if there is such a change) possibly happens at around 10 Myr. We note that 2MASS J1207334$-$393254b lacks {\sl Spitzer} photometry. More accurate data of L-type substellar members of young star clusters and associations (with ages below 10 Myr) are required for a precise characterization of the color dependency on gravity.

\begin{figure}
\includegraphics[angle=0,scale=0.47]{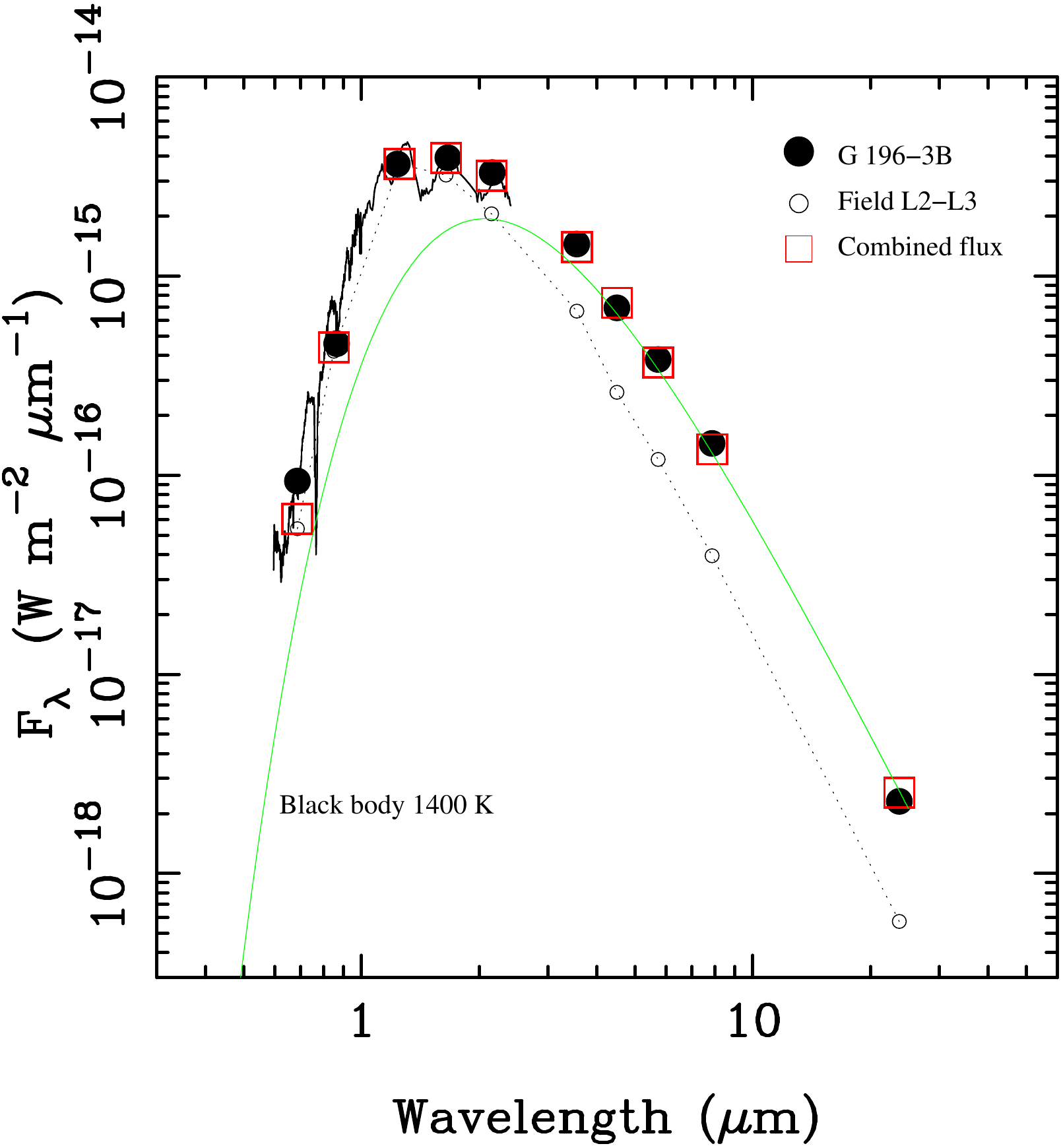}
\caption{The observed SED of G\,196--3B (solid line and filled dots) is reproduced by combining the flux emission of a field L2--L3 dwarf (dotted line and open circles) and a black body of single temperature, 1400 K, and $J$-band flux 4 times smaller than the dwarf. The combined SED (L2--L3 and black body) is plotted as open squares.
\label{bb}}
\end{figure}

The observed optical and infrared SED of \object{G\,196--3\,B} can be modeled by combining the emissions of an object with the same energy distribution than field L2--L3 dwarfs scaled to fit the $J$-band flux of the target and an additional source emitting like a single temperature black body. The L2--L3 dwarf component would mostly account for the visible up to the near-infrared fluxes, while the ``black body'' source would contribute notably to the mid-infrared fluxes. We found that a black body of 1400 K with a $J$-band flux four times fainter than that of the L2--L3 dwarf provides a reasonable fit to the thermal emission of the substellar object. Slightly cooler black bodies produce larger fluxes at red wavelengths while warmer ones yield insufficient emission. Figure~\ref{bb} illustrates our results. The agreement between the SED of \object{G\,196--3\,B} and the modeled SED is within 1--2-$\sigma$ uncertainty for the $I$ through [24] passbands, while the $R$-band is not well reproduced. No combination of L2--L3 dwarfs and observed SEDs of cooler field dwarfs (e.g., L2--L3\,$+$\,L7.5,  L2--L3\,$+$\,T1) provides a better match to \object{G\,196--3\,B}. The integrated luminosities of both the L2--L3 dwarf and the ``black body'' source are nearly equal even though their effective temperatures differ by about 500--700 K. If the black-body source is external to \object{G\,196--3\,B}, this would imply that the bolometric luminosity of the brown dwarf is to be corrected by about a factor of two (0.30 dex) towards lower luminosity.

We may speculate that \object{G\,196--3\,B} is comprised of an L2--L3 component and a cooler companion with a characteristic temperature of 1400 K and a $J$-band flux four times lower than the L2--L3 dwarf. High spatial resolution images (work in preparation) do not reveal \object{G\,196--3\,B} as a binary, so we can impose an upper limit of 0\farcs3 on the physical separation of both putative components. In the field, dwarfs of $\sim$2100 and $\sim$1400 K have bolometric luminosities differing by a factor of five. However, from our simple assumption we derived that both components show a luminosity ratio close to one, i.e., the cooler member should be 1.5 times larger in size than the warmer component. Substellar evolutionary models (Baraffe et al$.$ \cite{baraffe98}; Burrows et al$.$ \cite{burrows97}) predict that objects with masses around the deuterium-burning mass limit (i.e., 0.012--0.02 $M_\odot$) and ages below $\sim$100--200 Myr must burn deuterium. During this stage, there is a ``luminosity reversal'' because these objects become brighter than the immediately more massive brown dwarfs and at least as luminous as 0.025--0.030-$M_\odot$ bodies. Therefore, for an age younger than $\sim$100--200 Myr, \object{G\,196--3\,B} might be formed by two objects with masses of 0.012--0.02 and 0.02--0.03 $M_\odot$ and similar luminosity. There is an issue, however, with the predicted surface temperatures and object sizes. The same evolutionary models assign similar temperatures and a nearly constant radius to both the deuterium-burning and non-burning bodies of these masses. This is not consistent with our modeling of \object{G\,196--3\,B}'s SED, which suggests discrepant temperatures and sizes for the two putative members, undermining the probability of the multiplicity hypothesis.

Alternatively, the 1400-K black-body-like emission may result from an optically thick disk or envelope surrounding the central brown dwarf. Luhman et al$.$ \cite{luhman08}, Scholz \& Jayawardhana \cite{scholz08}, Zapatero Osorio et al$.$ \cite{osorio07}, and Caballero et al$.$ \cite{caballero07} showed that a large fraction of brown dwarfs and free-floating planetary-mass objects harbor disks at ages below 10 Myr. These disks may evolve into debris disks in a manner similar to the disks of stars. The presence of a (debris) disk is compatible with the young age of \object{G\,196--3\,B}. More than 40\%~of the stars with $\le$100 Myr display flux excesses at 24 $\mu$m due to the presence of debris disks (G\'asp\'ar et al$.$ \cite{gaspar09}). However, ``hot'' dust is rare in stars older than $\sim$10 Myr (e.g., Silvertone et al$.$ \cite{silverstone06}). Regarding \object{G\,196--3\,B}, the $\sim$1400 K effective temperature of the disk warm dust implies that the bulk of the observed material lies in a narrow belt $\sim$1 radius from the brown dwarf, making this scenario unlikely. 

One may think that the dense envelope has actually developed as a result of natural thermochemical processes in the upper, low-pressure atmospheric layers of \object{G\,196--3\,B}, which are producing dust grains small enough to be sustained in the low-gravity photosphere. These would be responsible for the near- to mid-infrared reddening. The 1400-K characteristic temperature might be related to the $T_{cr}$ parameter introduced by Tsuji \cite{tsuji02} in his model atmospheres, which essentially determines the thickness of the photospheric dust clouds or layers. Low values of $T_{cr}$ imply very dusty (optically thick) atmospheres.

Another possibility is that the warm dusty disk around  \object{G\,196--3\,B} may have appeared from ``recent'' and quite frequent stochastic events like episodes of high energy collisions among planetesimals (akin to the solar system's Late Heavy Bombardment and/or to the processes that are supposed to give rise to rocky planets, see review by Wyatt \cite{wyatt08}). In this scenario, the disk may be situated further away from the central object. This would suggest that even low-mass brown dwarfs are capable of forming planets through the same processes than more massive stars.

Any possibility previously discussed to account for the unusual red near- and mid-infrared colors of \object{G\,196--3\,B} can be also applied to the field young L dwarfs with related spectroscopic and photometric properties. Given the uncertainties associated to the models, we cannot firmly exclude any of the possibilities. However, a different metallicity seems to be the less likely explanation, while the effects on the output energy distribution driven by low gravity and enshrouded atmospheres and the presence of a warm dusty disk/envelope surrounding these objects remain as the most likely explanations. Follow-up radial velocity studies and polarimetric and high-resolution imaging will reveal whether these objects are multiple and/or harbor certain amounts of dust, which would help understand their nature.

\section{Conclusions}

We presented near-infrared, low-resolution spectrum (obtained with TNG/NICS instrument and Amici prism) and mid-infrared photometry ({\sl Spitzer}/IRAC and MIPS) of the brown dwarf \object{G\,196--3\,B}, which is a wide companion of the young, active M2.5-type star \object{G\,196--3\,A} (Rebolo et al$.$ \cite{rebolo98}). Using the near-infrared spectrum (0.8--2.4 $\mu$m), we confirmed an L3 spectral classification for the brown dwarf in agreement with the L3 typing derived from optical data by Cruz et al$.$ \cite{cruz09}; we also confirmed previously reported detection of spectroscopic features (``triangular'' shape of the $H$-band and strong VO$+$H$_2$O absorption) consistent with a low-gravity atmosphere and a young age. New optical and near-infrared images obtained on different occasions between 1998 and 2010, in addition to data available from public archives, allowed us to determine an accurate proper motion for each member of the binary and a projected separation and position angle of 15\farcs99\,$\pm$\,0\farcs06 and 209.3\,$\pm$\,0.3 deg, which have not changed in the last 12 yr (i.e., orbital motion below our astrometric detectability). This provides strong support for the gravitational link of the pair. We found that the kinematics of \object{G\,196--3} (adopting a probable distance interval of 15--30 pc) is consistent with the binary being a likely member of the Local Association, thus supporting the young age (20--300\,Myr) assumed for the system. A simple exercise of linearly back tracing the Galactic orbit of \object{G\,196--3} located the system relatively close to the past positions of the open clusters Collinder~65 (20--30 Myr) and $\alpha$~Persei (50--85 Myr), where \object{G\,196--3} may have originated. This implies that the age of the system could likely be 20--85 Myr (i.e., younger than the Pleiades star cluster). At this young age, \object{G\,196--3\,B} would have a mass of 0.012--0.25~$M_{\odot}$ and would be burning deuterium. 

From the {\sl Spitzer}/IRAC and MIPS photometry, we concluded that \object{G\,196--3\,A} spectral energy distribution (SED) covering 0.44--24 $\mu$m is purely photospheric, and that \object{G\,196--3\,B} shows infrared colors significantly redder than expected for its L3 spectral classification and much more similar to those of L7--L8 sources. The integration of the observed SEDs yielded a bolometric magnitude difference $\Delta m_{\rm bol}$\,=\,6.15\,$\pm$\,0.10 mag between the two objects. The SED of \object{G\,196--3\,B} appears overluminous longwards of 1.6 $\mu$m when compared to the average SED of field L2--L3 dwarfs. We discussed that a low-gravity atmosphere with enshrouded upper atmospheric layers and/or a warm dusty disk/envelope provides the most likely explanations, the two of them consistent with an age in the interval 20--85 Myr. \object{G\,196--3\,B} can be used as a benchmark object to understand L-type sources with similar photometric and spectroscopic properties in the field.

\acknowledgments

We are grateful to D. Montes for the discussion on object membership in stellar moving groups. We also appreciate the comments received by S. Mohanty (referee). This work is based in part on observations made with the {\sl Spitzer} Space Telescope, which is operated by the Jet Propulsion Laboratory, California Institute of Technology under a contract with NASA. Also based on new observations made with the Gran Telescopio Canarias (GTC), the Italian Telescopio Nazionale Galileo (TNG, operated by the Fundaci\'on Galileo Galilei of the Instituto Nazionale di Astrofisica), and the Isaac Newton Telescope (INT, operated by the Isaac Newton Group), instaled in the Spanish Observatorio del Roque de los Muchachos of the Instituto de Astrof\'\i sica de Canarias, in the island of La Palma. Also based on observatons taken with the 3.5-m telescope on the Calar Alto Spanish-German Observatory in Almer\'\i a, Spain. This research has benefitted from the SpeX Prism Spectral Libraries, maintained by Adam Burgasser at http://www.browndwarfs.org/spexprism. Funding for the Sloan Digital Sky Survey (SDSS) and SDSS-II has been provided by the Alfred P. Sloan Foundation, the Participating Institutions, the National Science Foundation, the U.S$.$ Department of Energy, the National Aeronautics and Space Administration, the Japanese Monbukagakusho, and the Max Planck Society, and the Higher Education Funding Council for England. The SDSS Web site is http://www.sdss.org/. This work is partly financed by the Spanish Ministry of Science through the project AYA2007--67458, and the CSIC I3 project 200940I010.

{\it Facilities:} \facility{Spitzer}, \facility{Telescopio Nazionale Galileo}, \facility{Gran Telescopio de Canarias}, \facility{Isaac Newton Telescope}, \facility{3.5-m Calar Alto}.

\clearpage

\clearpage

\begin{deluxetable}{lcc}
\tablecaption{Photometry of G\,196--3 \label{phot}}
\tablewidth{0pt}
\tablehead{ 
\colhead{ }  & \colhead{G\,196--3\,A} & \colhead{G\,196--3\,B}
}
\startdata
Spectral Type                  & M2.5               & L3\,$\pm$\,1 \\
$r$ (mag)\tablenotemark{a}     & \nodata            & 22.44\,$\pm$\,0.20 \\
$i$ (mag)\tablenotemark{a}     & \nodata            & 19.88\,$\pm$\,0.03 \\
$z$ (mag)\tablenotemark{a}     & \nodata            & 17.83\,$\pm$\,0.02 \\
$J$ (mag)\tablenotemark{b}     &  8.08\,$\pm$\,0.02 & 14.83\,$\pm$\,0.05 \\
$H$ (mag)\tablenotemark{b}     &  7.41\,$\pm$\,0.02 & 13.65\,$\pm$\,0.04 \\
$K_s$ (mag)\tablenotemark{b}   &  7.20\,$\pm$\,0.02 & 12.78\,$\pm$\,0.03 \\
$[3.6]$ (mag)                  &  7.03\,$\pm$\,0.01 & 11.66\,$\pm$\,0.02 \\
$[4.5]$ (mag)                  &  6.99\,$\pm$\,0.01 & 11.47\,$\pm$\,0.04 \\
$[5.8]$ (mag)                  &  6.94\,$\pm$\,0.01 & 11.10\,$\pm$\,0.06 \\
$[8.0]$ (mag)                  &  6.93\,$\pm$\,0.01 & 10.83\,$\pm$\,0.04 \\
$[24]$ (mag)                   &  6.83\,$\pm$\,0.05 & 10.55\,$\pm$\,0.10 \\
$m_{\rm bol}$ (mag)              &  9.85\,$\pm$\,0.10 & 16.00\,$\pm$\,0.10 \\
$T_{\rm eff}$ (K)                &  3480\,$\pm$\,90   & 1870\,$\pm$\,100   \\
$\mu_{\alpha}\,{\rm cos}\delta$ (mas\,yr$^{-1}$) & $-$142.3\,$\pm$\,0.6 & $-$144\,$\pm$\,2 \\
$\mu_{\delta}$ (mas\,yr$^{-1}$)                  & $-$197.0\,$\pm$\,0.9 & $-$190\,$\pm$\,20 \\
\enddata
\tablenotetext{a}{Sloan (DR7) photometry.}
\tablenotetext{b}{2MASS photometry.}
\end{deluxetable}


\begin{deluxetable}{lcrr}
\tablecaption{{\sl Spitzer}/MIPS photometry of field L dwarfs \label{mips}}
\tablewidth{0pt}
\tablehead{
\colhead{Object} & \colhead{SpT} & \colhead{$[24]$} & \colhead{AOR} \\
\colhead{       } & \colhead{ } & \colhead{(mag)} & \colhead{ } 
}
\startdata
\object{2MASS\,J1439284$+$192915}  & L1        & 10.29\,$\pm$\,0.15 & 4298240 \\
\object{DENIS\,J105847.9$-$154817} & L3        & 11.22\,$\pm$\,0.15 & 4299008 \\
\object{2MASS\,J0036159$+$182110}  & L3.5      &  9.65\,$\pm$\,0.15 & 4299776 \\
\object{DENIS\,J050124.1$-$001045} & L4$\gamma$& 10.78\,$\pm$\,0.15 & 22136576 \\
\object{2MASS\,J1507476$-$162738}  & L5        &  9.66\,$\pm$\,0.15 & 21906176 \\
\object{2MASS\,J0825196$+$211552}  & L7.5      & 10.31\,$\pm$\,0.12 & 21906688 \\
\enddata
\end{deluxetable}


\begin{deluxetable}{lll}
\tablecaption{Astrometric epochs for G\,196--3\,A \label{mpA}}
\tablewidth{0pt}
\tablehead{ 
\colhead{Date}  & \colhead{Catalog} & \colhead{Reference}
}
\startdata
1898 Apr 02 & AC2000.2                     & Urban et al$.$ \cite{urban98}  \\
1955 Jan 29 & USNO-A2.0 (POSSI Red)        & Monet et al$.$ \cite{monet98}  \\
1983 Feb 14 & GSC1.2 (Plate 01RU)          & Lasker et al$.$ \cite{lasker88}\\
1984 Nov 20 & GSC1.2 (Plate 01RP)          & Lasker et al$.$ \cite{lasker88}\\
1989 Dec 06 & POSS II Red\tablenotemark{a} & This paper \\
1991 Jul 10 & GSC2.2 \& 2.3                & STScI 2001, 2006 \\
1992 Feb 05 & POSS II Blue\tablenotemark{a}& This paper \\
1999 Jan 19 & 2MASS                        & Skrutskie et al$.$ \cite{skrutskie06}  \\
1999 Feb 24 & POSS II IR\tablenotemark{a}  & This paper \\
2002 Feb 01 & SDSS DR7                     & Abazajian et al$.$ \cite{abazajian07} \\
2005 Nov 26 & {\sl Spitzer}/IRAC\tablenotemark{b} & This paper \\
2010 Jan 30 & 3.5-m CAHA/Omega-2000 & This paper \\
\enddata
\tablenotetext{a}{Astrometry obtained from the SuperCOSMOS digitalization of the Palomar Observatory Sky Survey II red, blue, and infrared photographic plates (Hambly et al$.$ \cite{hambly01}). We used common IRAF routines to measure the photocentroid of G\,196--3\,A.}
\tablenotetext{b}{Astrometry provided by the IRAC pipeline indicated in the text. We used common IRAF routines to measure the photocentroid of G\,196--3\,A.}
\end{deluxetable}

\begin{deluxetable}{lll}
\tablecaption{Astrometric epochs for G\,196--3\,B \label{mpB}}
\tablewidth{0pt}
\tablehead{ 
\colhead{Date}  & \colhead{Images} & \colhead{Reference}
}
\startdata
1998 Feb 16 & NOT/ALFOSC & Rebolo et al$.$ \cite{rebolo98} \\
1998 Jun 03 & NOT/HiRAC  & Rebolo et al$.$ \cite{rebolo98} \\
2008 May 07 & GTC/ASG    & This paper \\
2008 Nov 28 & INT/WFC    & This paper \\
2010 Jan 30 & 3.5-m CAHA/Omega-2000 & This paper \\
\enddata
\end{deluxetable}

\begin{deluxetable}{rcccc}
\tablecaption{Distance estimates for G\,196--3 \label{dist}}
\tablewidth{0pt}
\tablehead{
\colhead{Age} & \colhead{G\,196--3\,A} & \colhead{G\,196--3\,A} &
                \colhead{G\,196--3\,B} & \colhead{G\,196--3\,B}\\
\colhead{}    & \colhead{single} & \colhead{double} & 
                    \colhead{single} & \colhead{double}  \\
\colhead{(Myr)}   & \colhead{      } & \colhead{      } & 
                    \colhead{      } & \colhead{      }  \\
}
\startdata
Field & 15 & 22 & 18      & 25 \\
120   & 19 & 27 & 20      & 28 \\
50    & 29 & 41 & \nodata & \nodata \\
3     & 33 & 46 & 36      & 51 \\
\enddata
\tablecomments{Distance estimates are in pc and have an associated uncertainty of 10\%.}
\end{deluxetable}

\end{document}